\documentclass[aip, reprint]{revtex4-1}

\usepackage{graphicx}
\usepackage{amsmath}
\usepackage{hyperref}

\newcommand{\HFM}{(H$_2$)$_4$CH$_4$}

\begin{document}
%%%%%%%%%%%%%%%%%%%%%%%%%%%%%%%%%%%%%%%%%%%%%%%%%%%%%%%%%%%%%%%%%%%%%%%%

%%%%%%%%%%%%%%%%%%%%%%%%%%%%%%%%%%%%%%%%%%%%%%%%%%%%%%%%%%%%%%%%%%%%%%%%
\title{H4-Alkanes: A new class of hydrogen storage material?}

\author{D. Harrison}
\affiliation{Department of Physics, Wake Forest University,
Winston-Salem, NC 27109, USA.}

\author{E. Welchman}
\affiliation{Department of Physics, Wake Forest University,
Winston-Salem, NC 27109, USA.}

\author{T. Thonhauser}
\email{thonhauser@wfu.edu}
\affiliation{Department of Physics, Wake Forest University,
Winston-Salem, NC 27109, USA.}

\date{\today}

\begin{abstract}
The methane-based material \HFM, also called H4M for short, is in
essence a methane molecule with 4 physisorbed H$_2$ molecules. While H4M
has exceptionally high hydrogen storage densities when it forms a
molecular solid, unfortunately, this solid is only stable at impractically
high pressures and/or low temperatures.  To overcome this limitation, we
show through simulations that longer alkanes (methane is the
shortest alkane) also form stable structures that still physisorb 4
H$_2$ molecules per carbon atom; we call those structures H4-alkanes. We
further show via molecular dynamics simulations that the stability field
of molecular solids formed from H4-alkanes increases remarkably with
chain length compared to H4M, just as it does for regular alkanes. From
our simulations of H4-alkanes with lengths 1, 4, 10, and 20, we see that
e.g.\ for the 20-carbon the stability field is doubled at higher
pressures. While even longer chains show only insignificant
improvements, we discuss various other options to stabilize H4-alkanes
more.  Our proof-of-principle results lay the groundwork to show
that H4-alkanes can become viable hydrogen storage materials.
\end{abstract}

\maketitle
%%%%%%%%%%%%%%%%%%%%%%%%%%%%%%%%%%%%%%%%%%%%%%%%%%%%%%%%%%%%%%%%%%%%%%%%

%%%%%%%%%%%%%%%%%%%%%%%%%%%%%%%%%%%%%%%%%%%%%%%%%%%%%%%%%%%%%%%%%%%%%%%%
\section{Introduction}\label{sec:introduction}
%%%%%%%%%%%%%%%%%%%%%%%%%%%%%%%%%%%%%%%%%%%%%%%%%%%%%%%%%%%%%%%%%%%%%%%%

\begin{figure}[t]
\includegraphics[width=0.55\columnwidth]{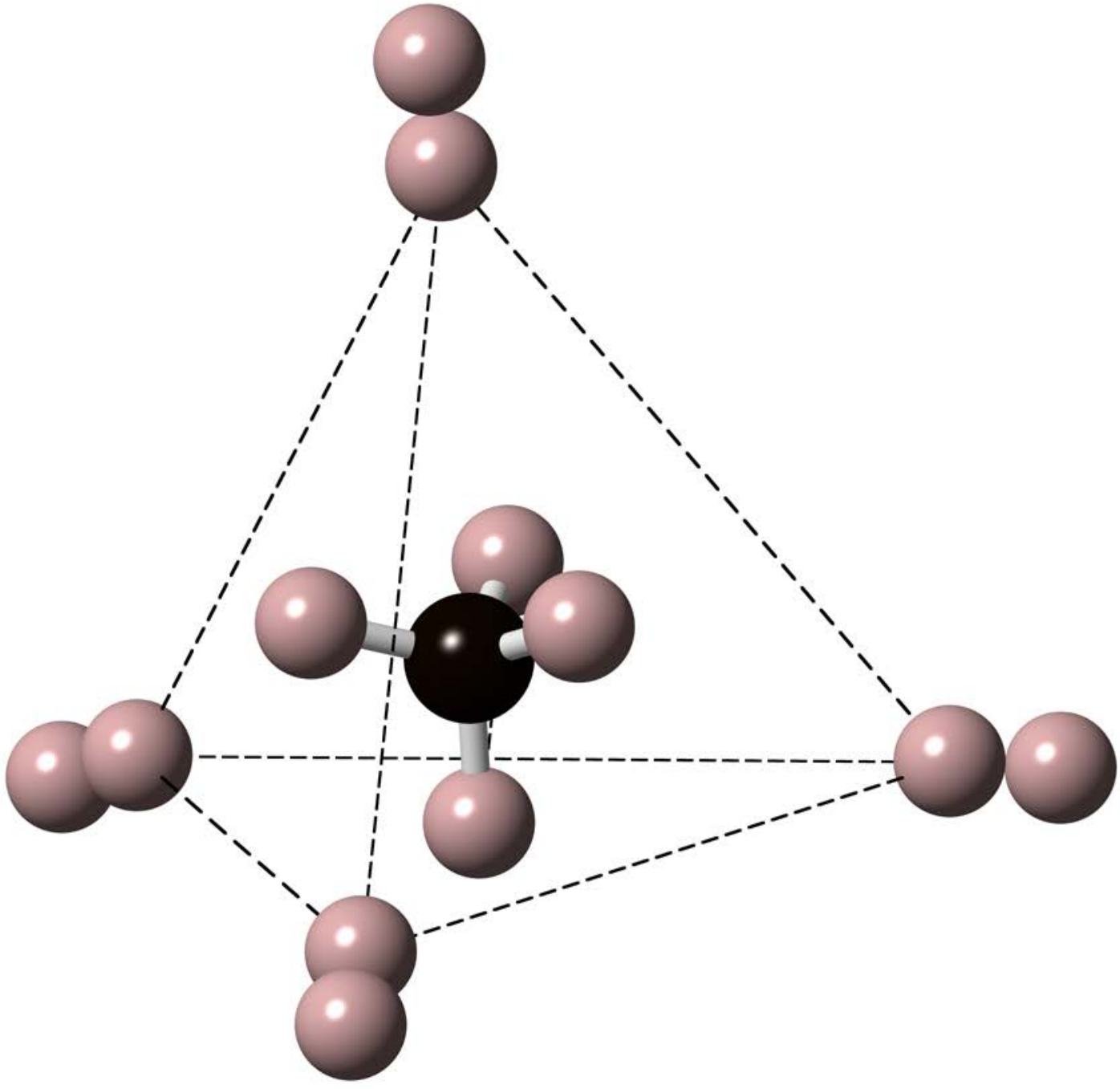}
\caption{\label{fig:H4M_molecule}A single \HFM\ molecule. This material
is also called H4M, as four hydrogen molecules are bound to one
methane.}
\end{figure}

In the search for a clean and renewable replacement for fossil fuels,
using hydrogen as an energy carrier represents the ideal
solution.\cite{Kunowsky_2013:material_demands,
Durbin_2013:review_hydrogen, Crabtree_2004:hydrogen_economy}
Unfortunately, hydrogen is not practical for mobile applications in its
natural state because of its low volumetric density. Many potential
approaches and materials for solving this problem have been
explored,\cite{Graetz_2009:new_approaches,
Harrison_2015:materials_hydrogen} but (H$_2$)$_4$CH$_4$---also known as
H4M---demonstrates the largest volumetric and gravimetric densities of
any known hydrogen storage material;\cite{Mao_2007:clathrate_hydrates,
Li_2012:theoretical_study} its 50.2~mass\% (33.3~mass\% without the
hydrogen in CH$_4$) and 0.15~kg~H$_2$/L far exceed the ultimate
gravimetric and volumetric density targets for hydrogen storage
materials set by the Department of Energy
(DOE).\cite{Yang_2010:high_capacity, DOE_Targets_Onboard_2009, Ahluwalia_2015:sorbent_material} H4M was
first discovered almost two decades
ago\cite{Somayazulu_1996:high-pressure_compounds} and contains four
H$_2$ molecules bound to a CH$_4$ methane molecule by van der Waals
interactions, see Fig.~\ref{fig:H4M_molecule}. Unfortunately, H4M
requires extreme conditions to remain stable (65~K at ambient pressure
or 5--6~GPa at ambient
temperature)\cite{Mao_2005:pressure-temperature_stability,
Struzhkin_2007:hydrogen_storage} and despite its great promise has seen
little study beyond property characterization. Due to the narrow
stability range of H4M, any attempt to harness its high storage
densities requires further stabilization. To this end, previous work has
attempted to use external agents such as metal organic frameworks,
carbon nanotubes, boron nitride nanotubes, graphite, or hexagonal boron
nitride.\cite{Li_2012:theoretical_study, Wang_2015:ab_initio} Although
relatively little work has been done on H4M specifically, other relevant
works include investigations of high-pressure systems and the extensive
work done on clathrates, a class of materials to which H4M is related.
\cite{Hemley_2000:effects_high, Song_2013:new_perspectives,
Lee_2005:tuning_clathrate, Florusse_2004:stable_low-pressure, Wang_2009:high_pressure, Gupta_2008:measurements_methane}

\begin{figure}[t]
\includegraphics[width=0.9\columnwidth]{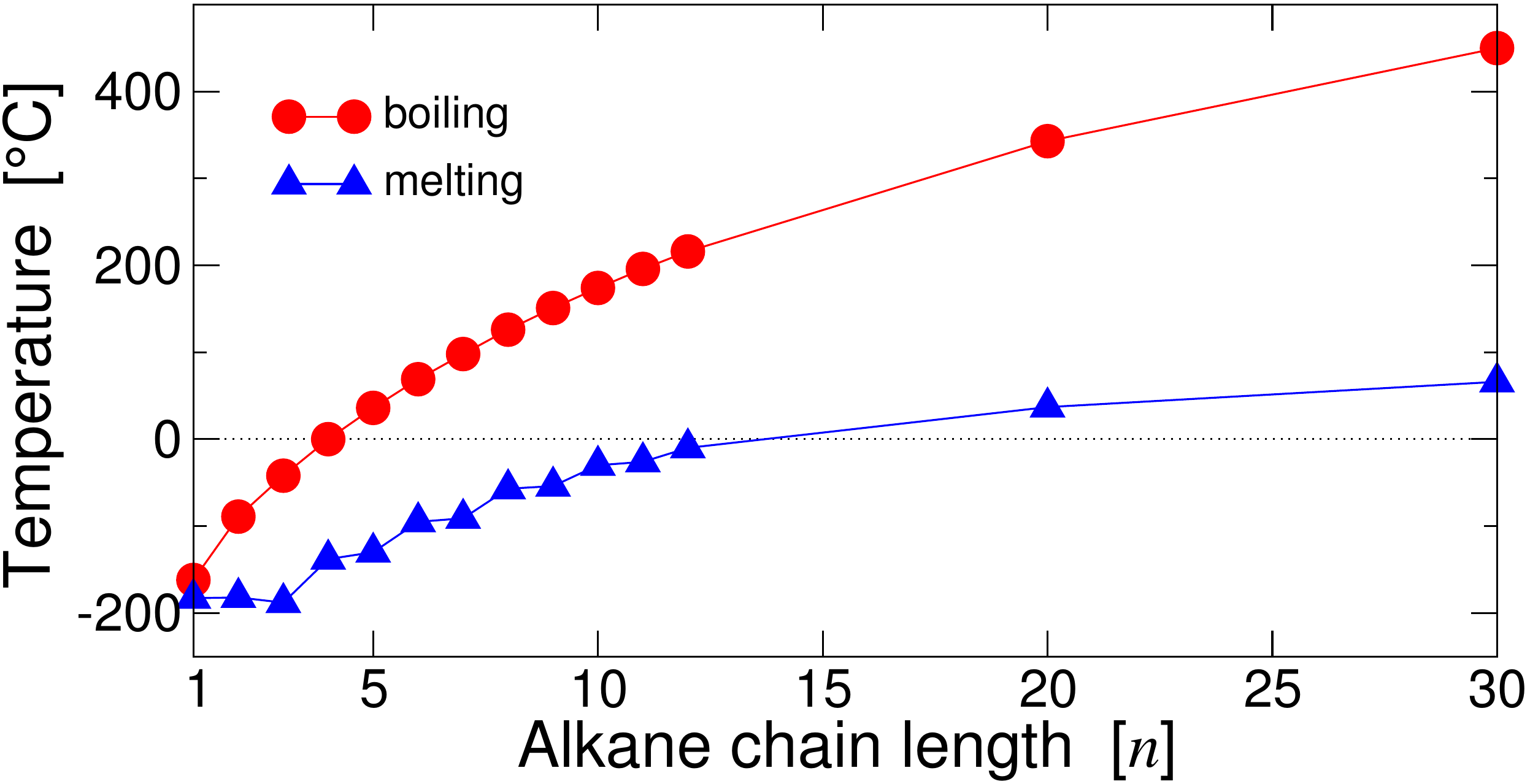}
\caption{\label{fig:melting}Melting and boiling temperatures for alkanes
as a function of chain length $n$. Data taken from standard organic
chemistry textbook.\cite{Loudon_1995:organic_chemistry}}
\end{figure}

Rather than try to improve the stability of H4M directly, we instead use
H4M as the inspiration to study a new class of materials. Alkanes are
hydrocarbon chains of the form C$_n$H$_{2n+2}$.  Of particular interest
is their increase in melting and boiling point with chain length $n$, as
seen in Fig.~\ref{fig:melting}. For example, as chain length increases
from $n=1$ to $n=6$, the melting and boiling temperatures increase by
over 100 and 200~$^\circ$C,
respectively.\cite{Loudon_1995:organic_chemistry} This increase in the
melting/boiling temperatures is due to the additional long-range
interactions among the longer chains.  Because H4M is essentially an
alkane chain with $n=1$ (i.e.\ methane) and 4 physisorbed H$_2$
molecules, we hypothesize that if we physisorb 4~H$_2$ per carbon atom
to a longer alkane chain, it would form a stable compound with a higher
melting point than H4M. This results in a new class of materials of the
form (H$_2$)$_{4n}$C$_n$H$_{2n+2}$, which we call \emph{H4-alkanes}.  In
the following, we show that such materials can be stable and that the
stability of these materials increases with chain length, analogous to
how the stability of regular alkane chains increases with length. To
this end, we calculate the phase diagrams of several H4-alkanes of
various lengths. In particular, we determine the melting temperatures of
chains with lengths $n=1$, 4, 10, and $20$ for pressures ranging from
0~GPa to 6~GPa in increments of 1~GPa. These chain lengths were chosen
because they are representatives of gaseous ($n=4$), liquid ($n=10$),
and solid alkanes ($n=20$) under ambient conditions; even chain lengths
$n$ were chosen because they show a larger increase of stability going
from $n-1$ to $n$ compared to odd ones (see e.g.\ the step-structure in
the blue line in Fig.~\ref{fig:melting}). We will refer to H4-alkanes of
length $n$ as ``$n$-carbon,'' except for the case of $n=1$, which we will
continue to call H4M.

%%%%%%%%%%%%%%%%%%%%%%%%%%%%%%%%%%%%%%%%%%%%%%%%%%%%%%%%%%%%%%%%%%%%%%%%
\section{Computational Details}
%%%%%%%%%%%%%%%%%%%%%%%%%%%%%%%%%%%%%%%%%%%%%%%%%%%%%%%%%%%%%%%%%%%%%%%%

%%%%%%%%%%%%%%%%%%%%%%%%%%%%%%%%%%%%%%%%%%%%%%%%%%%%%%%%%%%%%%%%%%%%%%%%
\subsection{Structure Searches}
%%%%%%%%%%%%%%%%%%%%%%%%%%%%%%%%%%%%%%%%%%%%%%%%%%%%%%%%%%%%%%%%%%%%%%%%

Structure searches for the various H4-alkanes with $n=1$, 4, 10, and
$20$ were performed with the structure search program Universal
Structure Predictor: Evolutionary Xtallography
(\textsc{Uspex})\cite{Oganov_2006:crystal_structure,
Oganov_2011:how_evolutionary, Lyakhov_2013:new_developments,
Zhu_2012:constrained_evolutionary} in conjunction with the program 
Large-Scale Atomic/Molecular Massively Parallel Simulator (\textsc{Lammps}).\cite{Plimpton_1995:fast_parallel}
The potential used was the Adaptive Intermolecular Reactive Empirical Bond
Order (\textsc{Airebo})
potential,\cite{Brenner_2002:second-generation_reactive,
Stuart_2000:reactive_potential} which is specifically parameterized for
hydrocarbons; a value of 3.0 was used for the Lennard--Jones sigma scale
factor. Note that the \textsc{Airebo} potential does not include three-body
dispersive interactions, which can have small but noticeable
effects.\cite{Anatole-von-Lilienfeld_2010:two-_three-body}
\textsc{Uspex} was
configured to run a three-dimensional molecular structure search---with the
two constituent molecules being the alkane and hydrogen molecule---with a
population size of 20 at each
generation.\cite{Zhu_2012:constrained_evolutionary} Structures kept from previous generations were re-relaxed and a stopping criteria of 6
generations was used; that is, if the same structure had the lowest energy for 6
consecutive generations, the structure search was ended. Due to their smaller
size, 8 and 4 alkane molecules (and the corresponding number of hydrogen
molecules) were used for H4M and the 4-carbon, respectively, while for the 10-carbon
and 20-carbon 2 alkane molecules were used per unit cell. For each generation,
the structures produced by \textsc{Uspex} were relaxed in \textsc{Lammps},
first by running a short molecular dynamics (MD) simulation at 10--50~K (increasing linearly) and 6~GPa for
0.5~ps before using the conjugate gradient method (again with the simulation cell at a pressure of 6~GPa).

%%%%%%%%%%%%%%%%%%%%%%%%%%%%%%%%%%%%%%%%%%%%%%%%%%%%%%%%%%%%%%%%%%%%%%%%
\subsection{Molecular Dynamics Modeling}
%%%%%%%%%%%%%%%%%%%%%%%%%%%%%%%%%%%%%%%%%%%%%%%%%%%%%%%%%%%%%%%%%%%%%%%%

The best (i.e.\ lowest energy) structures obtained from the structure
search were then used as the starting structure for our MD simulations, again using \textsc{Lammps}
in conjunction with the 
\textsc{Airebo}
potential. A time-step of 0.5~fs was used. For each H4-alkane, a
supercell was created such that the length of each unit cell vector was
at least 19~\AA, resulting in at least 2200 atoms per unit cell. Many MD simulations were then performed on the system at a variety of
temperatures and pressures, with pressure ranging from 1~atm to
60,000~atm in increments of 10,000~atm (i.e. $\sim$0~GPa to 6~GPa in
increments of 1~GPa) and temperatures being probed in increments of
10~K until a clear criteria of melting was fulfilled. Every simulation
was run for a total of 11~ns, with the first nanosecond being used for
equilibration and the remaining 10 for analyzing melting criteria.

%%%%%%%%%%%%%%%%%%%%%%%%%%%%%%%%%%%%%%%%%%%%%%%%%%%%%%%%%%%%%%%%%%%%%%%%
\subsection{Criteria for Melting}
%%%%%%%%%%%%%%%%%%%%%%%%%%%%%%%%%%%%%%%%%%%%%%%%%%%%%%%%%%%%%%%%%%%%%%%%

\begin{figure}[t]
\includegraphics[width=\columnwidth]{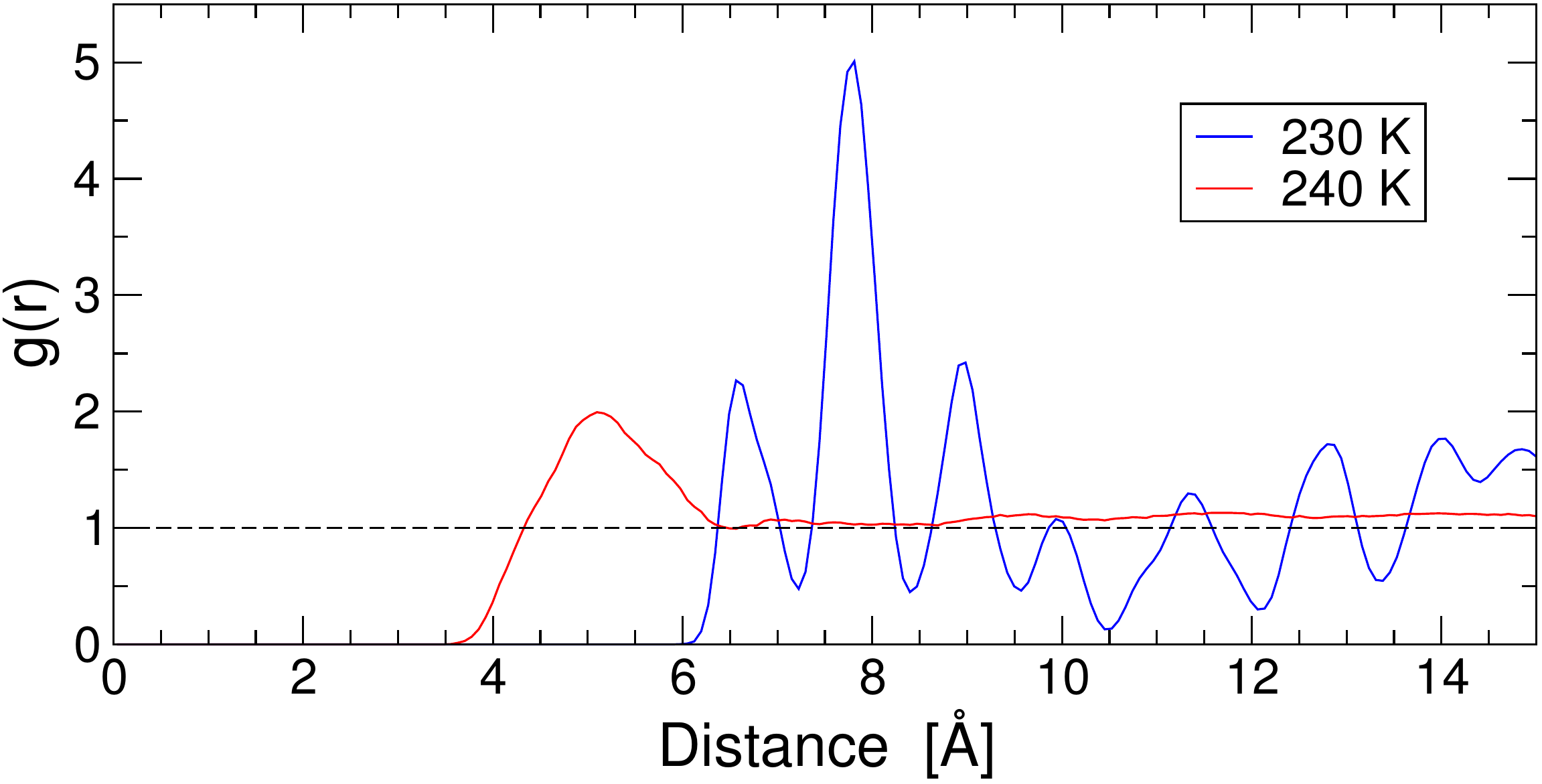}
\caption{\label{fig:radial}Alkane-alkane center-of-mass radial distribution function
$g(r)$ of the 4-carbon at 6~GPa and 230~K and 240~K. While the material at 240~K
exhibits a complete loss of long-range order at
7~\AA\ and can thus be considered molten, at 230~K there is order out to 15~\AA\ and beyond, indicating a solid phase. The units of
the y-axis are normalized so that 1 corresponds to the average density.}
\end{figure}

\begin{figure}[t]
\includegraphics[width=\columnwidth]{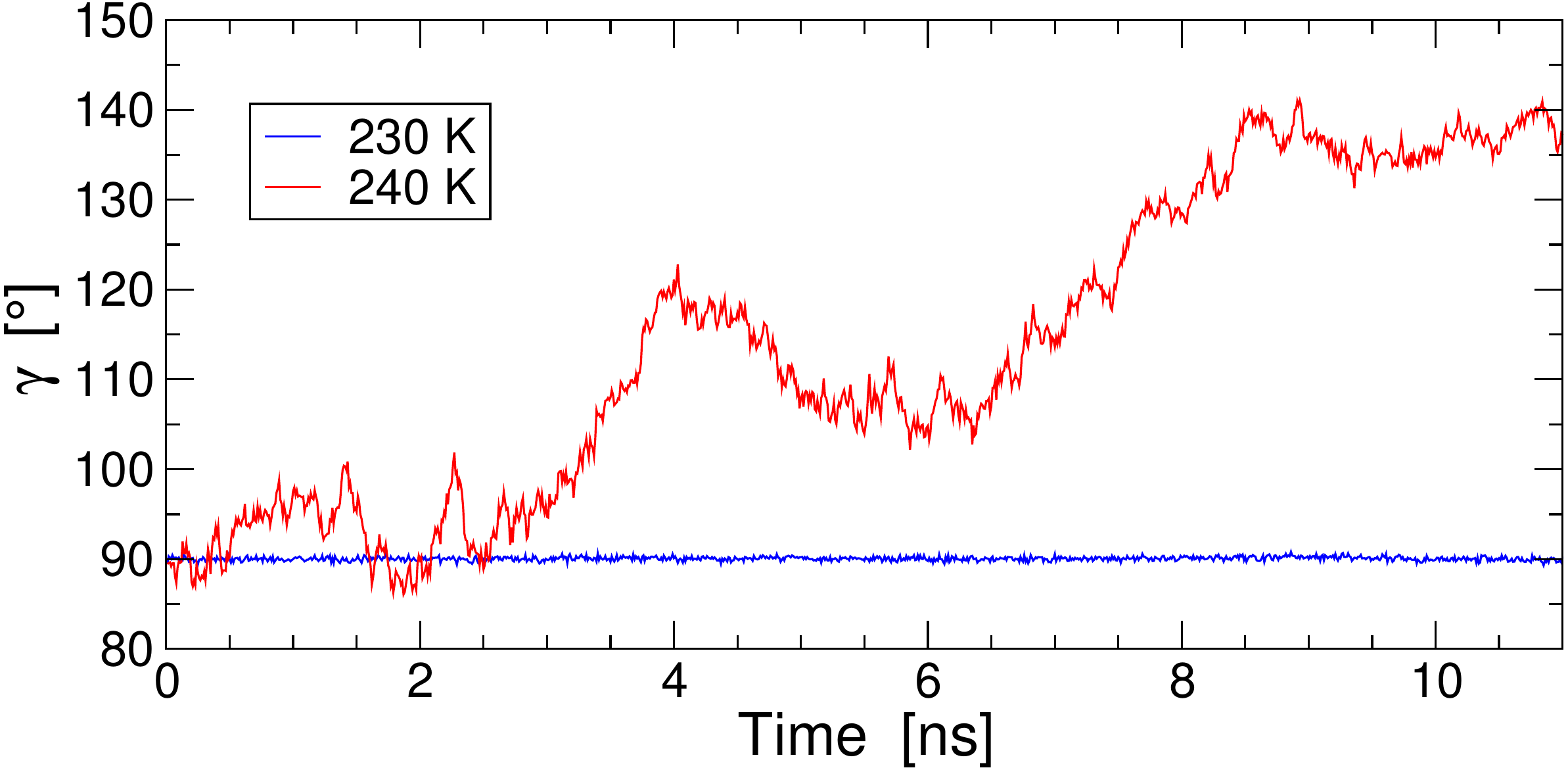}
\caption{\label{fig:angle}Lattice angle $\gamma$ of the 4-carbon over a full MD run
at 6~GPa. A clear transition is visible, introducing a significant deviation,
when the temperature changes
from 230~K to 240~K. The occurrence of this distinct feature coincides exactly
with the change in the radial distribution function in Fig.~\ref{fig:radial}
and thus provides a convenient metric for determining whether a particular
H4-alkane is molten or solid.}
\end{figure}

Rather than using the radial distribution function---a
standard method for identifying melting in a material---we designed a
simple, non-graphical order parameter to streamline the process of
identifying melting in a large number of simulations. We used the
following criteria: if the standard deviation in any of the lattice
angles $\alpha$, $\beta$, or $\gamma$ was greater than 5$^{\circ}$ (with
data taken over the last 10 nanoseconds of our MD run), it
was considered to be a liquid.  The justification for this criteria is
that liquids have zero shear modulus, and the shear modulus is
essentially a shear stress divided by the resultant change in angle, so
we expect a large increase in the standard deviation of the lattice
angles after the material melts. We found this criteria to agree very
well with the loss of long-range order in the alkane-alkane center-of-mass radial distribution plots
for all of our H4-alkanes. As an example, the standard
deviations of ($\alpha$, $\beta$, $\gamma$) in degrees for the 4-carbon at 6~GPa and 230~K
and 240~K (with radial distribution functions shown in
Fig.~\ref{fig:radial}) were (0.295, 0.306, 0.236) and (12.280,
14.325, 16.223), respectively. Compared to Fig.~\ref{fig:radial}, we can
see that the loss of structure going from 230~K to 240~K corresponds
very well with a sharp increase in the variance of the lattice angles.
We explicitly show how the lattice angle behavior changes upon melting in
Fig.~\ref{fig:angle}, where we plot $\gamma$ over the full MD run. We
see this same degree of correspondence between radial distribution
function and angle variance at all of the pressures we have studied (0
to 6~GPa). For a few selected cases
we have also studied the energy vs.\ temperature behavior and saw `jumps'
at the phase transitions. However, the melting temperatures derived from
this approach coincide exactly with the temperatures found by our method of
analyzing the deviation in lattice angles, which we find computationally
easier to monitor and automate.

Another way of determining melting would be with a two-phase
calculation, where one phase is in the solid state and the other phase
is in the liquid state. Below and above the melting temperature, one of
the phases grows while the other shrinks. This process is not applicable
here, as the unit cells are already very large for the longer chains and
alkanes become very ``waxy'' structures as the chains get longer,
resulting in a slow moving of the phase boundary. In addition, the
structures of the liquid state of H4-alkane are not known either and
would require structure searches at elevated temperatures, for which no
clear path exists.

%%%%%%%%%%%%%%%%%%%%%%%%%%%%%%%%%%%%%%%%%%%%%%%%%%%%%%%%%%%%%%%%%%%%%%%%
\section{Results and Discussion}
%%%%%%%%%%%%%%%%%%%%%%%%%%%%%%%%%%%%%%%%%%%%%%%%%%%%%%%%%%%%%%%%%%%%%%%%

%%%%%%%%%%%%%%%%%%%%%%%%%%%%%%%%%%%%%%%%%%%%%%%%%%%%%%%%%%%%%%%%%%%%%%%%
\subsection{H4-Alkane Structures}
%%%%%%%%%%%%%%%%%%%%%%%%%%%%%%%%%%%%%%%%%%%%%%%%%%%%%%%%%%%%%%%%%%%%%%%%

\begin{figure*}
\hspace*{\fill}
a)\includegraphics[width=0.5\columnwidth]{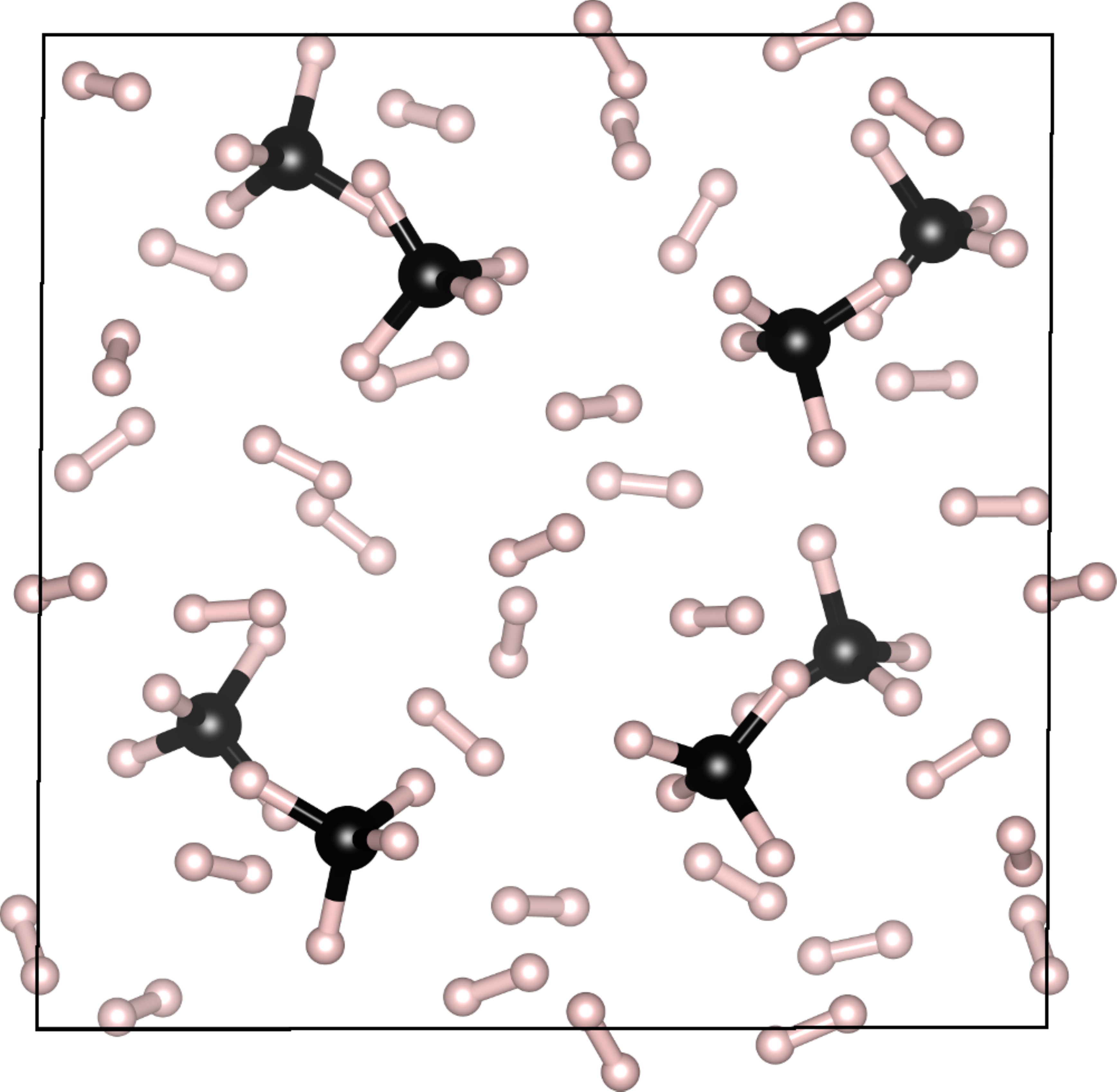}\hfill\hfill
b)\includegraphics[width=0.75\columnwidth]{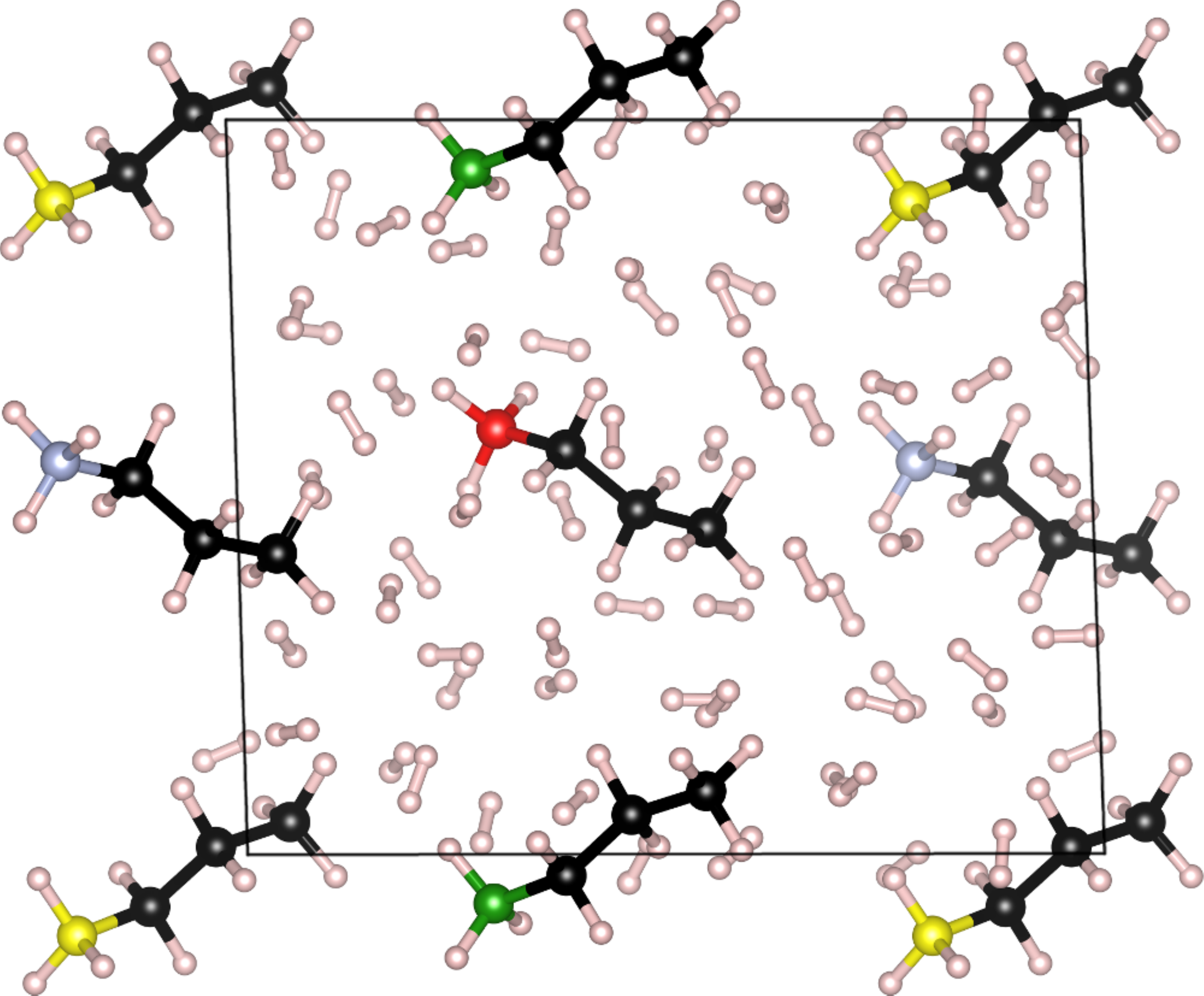}\hfill\mbox{}\\[5ex]
c)\includegraphics[width=1\columnwidth]{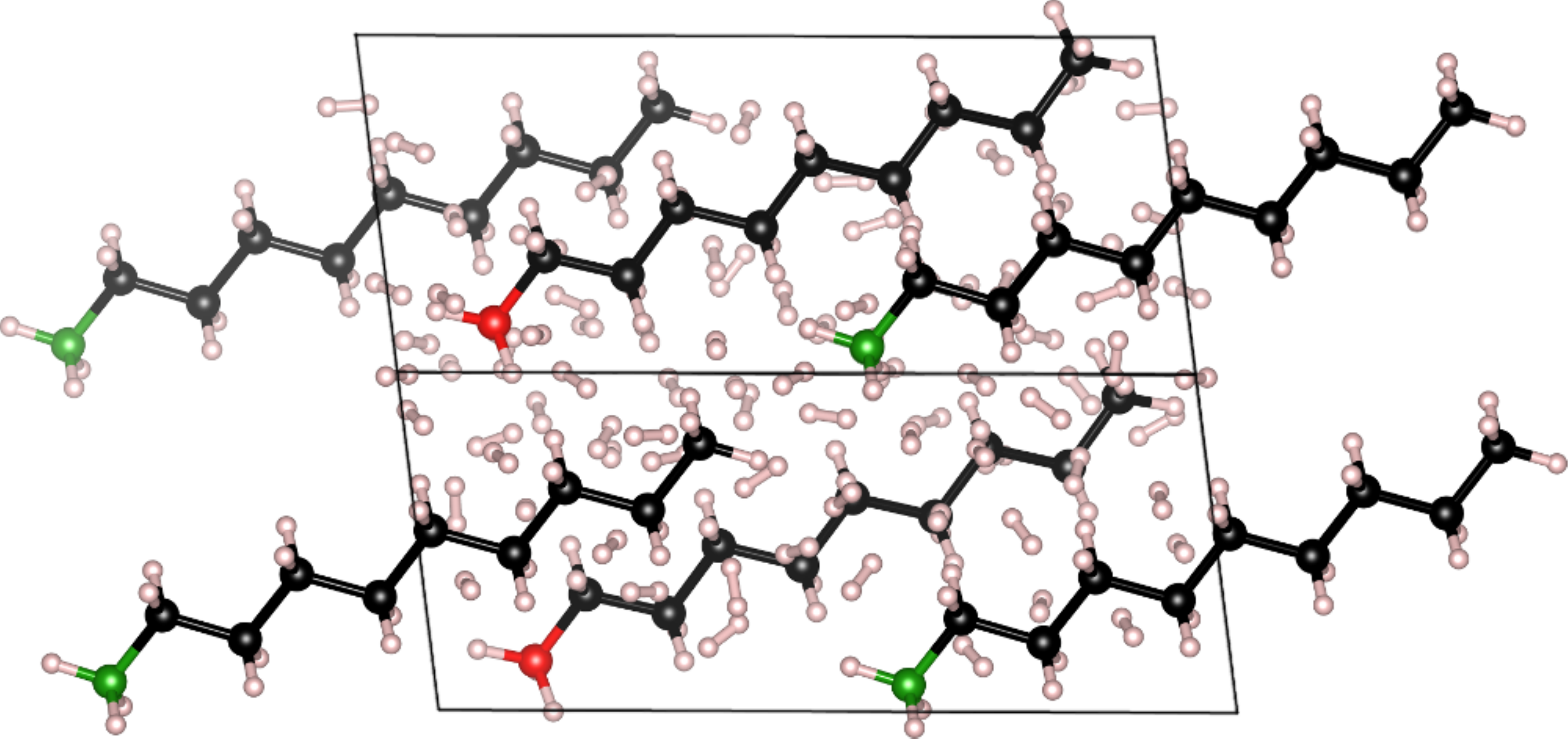}\hfill
d)\includegraphics[width=1\columnwidth]{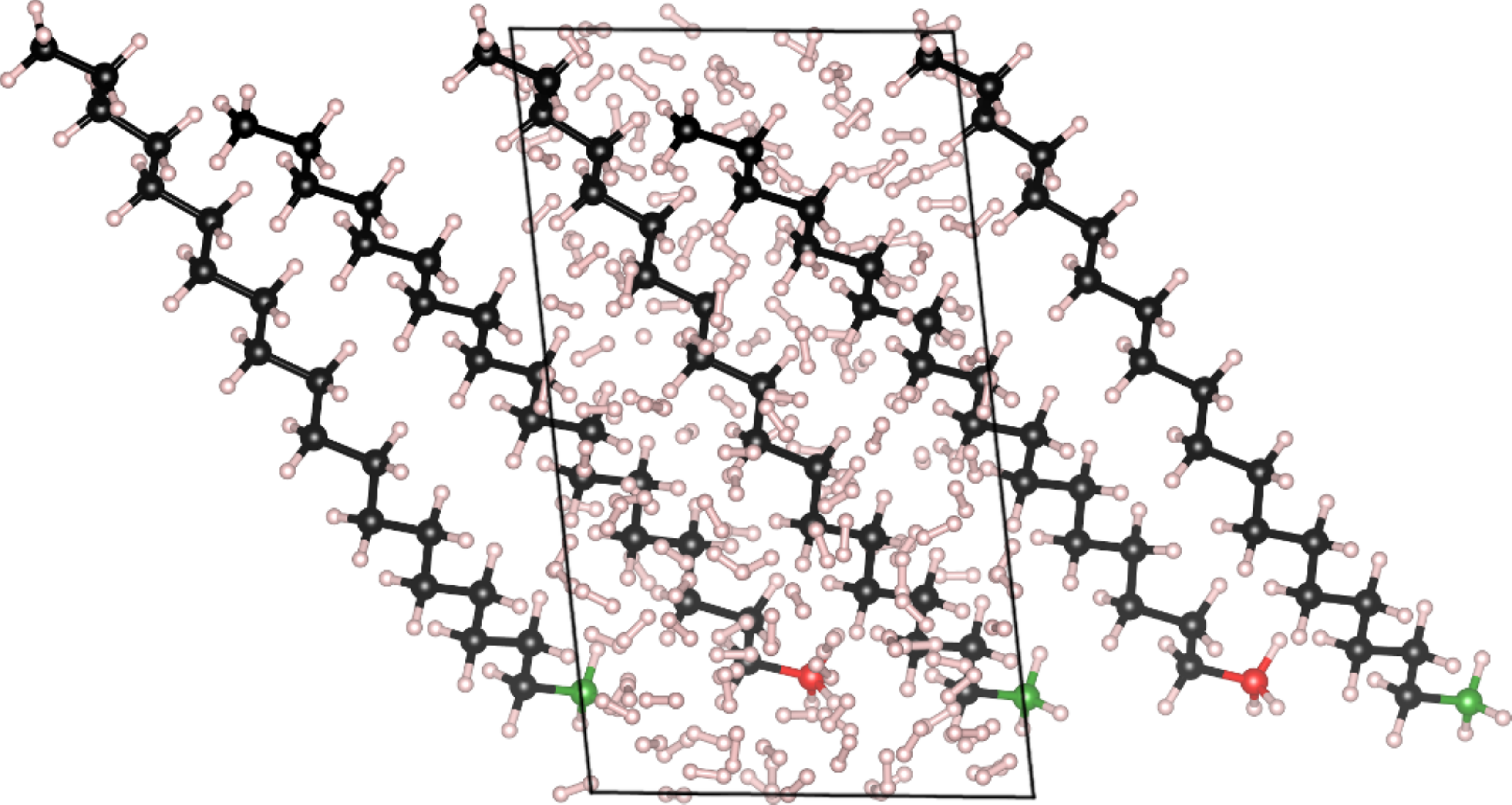}
\caption{\label{fig:structures}Structures found for a) H4M, b) 4-carbon, c) 10-carbon,
and d) 20-carbon. Certain carbon atoms have been color-coded so that it is clear
which chains are distinct and which are periodic images.}
\end{figure*}

Our goal in performing a structure search for H4M or any of the other
H4-alkanes was not foremost to find the true ground-state structures
and global minimum, but to have a consistent method for generating a
reasonable starting structure for our MD simulations.
Nonetheless, we briefly discuss here the structures found for H4M and
the other H4-alkanes, shown in Fig.~\ref{fig:structures}.  Note that simply putting H4-alkanes more or less
arbitrarily into a large unit cell results in a lot of chain-breaking,
which the structure search avoids.

Although the true structure of H4M remains unknown, previous
investigations have suggested highly ordered structures (e.g.\
body-centered tetragonal or body-centered orthogonal).\cite{Somayazulu_1996:high-pressure_compounds,
Li_2012:theoretical_study, High_capacity_molecular_2009} In contrast, although
we find an approximately orthorhombic unit cell, it is not body-centered. Note that
Ref.~[\onlinecite{High_capacity_molecular_2009}] screened many different
randomly generated structures using classical force fields, whereas
Ref.~[\onlinecite{Li_2012:theoretical_study}] did not generate random
structures, but instead compared several high-symmetry structures using
DFT. In contrast, we used an evolutionary random structure search
algorithm in conjunction with \textsc{Lammps}. Comparing the energy of our H4M structure to a body-centered orthogonal structure (such as suggested by Ref.~[\onlinecite{High_capacity_molecular_2009}]) at a pressure
of 6~GPa upon relaxation using \textsc{Lammps}, we find the structures to be 
nearly isoenergetic, with an energy difference of less than 4~meV/atom. 

For the other H4-alkanes we find that, as the chain length increases,
there is more order among the chains. The 4-carbon structures seem to have aligned alkane
slabs with alternating tilt, whereas in the 10-carbon structure the alkane chains are pointed in the same direction,
although with an offset to one another. The 20-carbon structure is most ordered,
with the alkane chains mostly in alignment with one another, forming a rhombus.

%%%%%%%%%%%%%%%%%%%%%%%%%%%%%%%%%%%%%%%%%%%%%%%%%%%%%%%%%%%%%%%%%%%%%%%%
\subsection{Comparison with Experiment}
%%%%%%%%%%%%%%%%%%%%%%%%%%%%%%%%%%%%%%%%%%%%%%%%%%%%%%%%%%%%%%%%%%%%%%%%

\begin{figure}[t]
\includegraphics[width=\columnwidth]{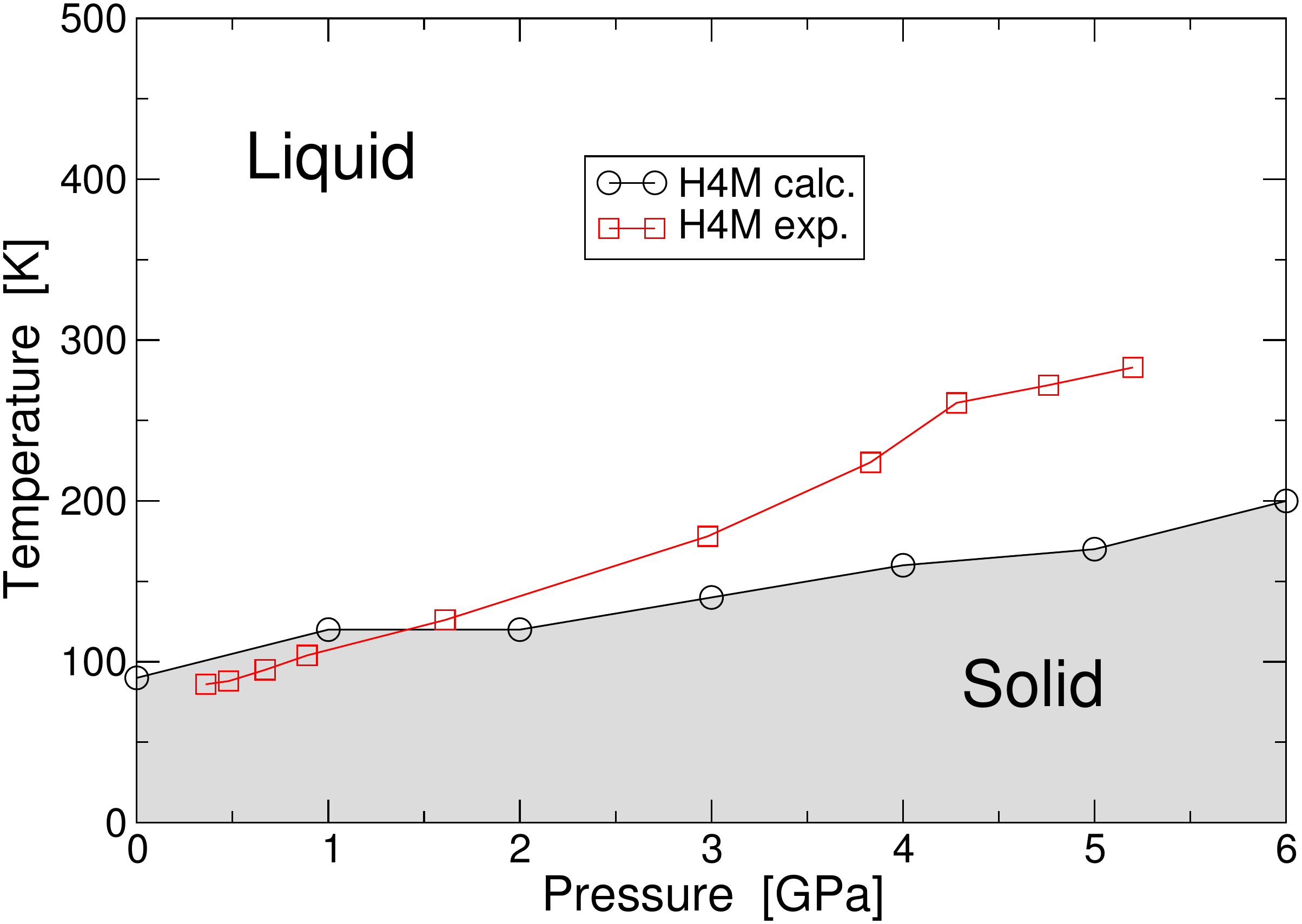}
\caption{\label{fig:h4m}Calculated phase diagram for H4M compared to
experimental data taken from Mao et
al.\cite{Mao_2005:pressure-temperature_stability,
Struzhkin_2007:hydrogen_storage} The experimental melting curve is
defined by the points where H4M solidified upon temperature decrease.}
\end{figure}

To test our approach, we first perform a structure search and MD
simulations for H4M to find its phase diagram, which can be compared
with experiment.  In Fig.~\ref{fig:h4m}, we find near exact agreement of
the phase diagram for lower pressures, although our results increasingly
under-predict the melting temperature compared to experiment as the
pressure increases. This error is likely due to the \textsc{Airebo} potential
being optimized for lower pressures. Even so, the error is still very
reasonable for melting temperatures calculated with empirical force
fields, especially considering we are more interested in trends, i.e.\
\emph{differences}, in melting temperatures as chain length increases,
rather than their absolute values.

%%%%%%%%%%%%%%%%%%%%%%%%%%%%%%%%%%%%%%%%%%%%%%%%%%%%%%%%%%%%%%%%%%%%%%%%
\subsection{Phase Diagram of H4-Alkanes}
%%%%%%%%%%%%%%%%%%%%%%%%%%%%%%%%%%%%%%%%%%%%%%%%%%%%%%%%%%%%%%%%%%%%%%%%

\begin{figure}[t]
\includegraphics[width=\columnwidth]{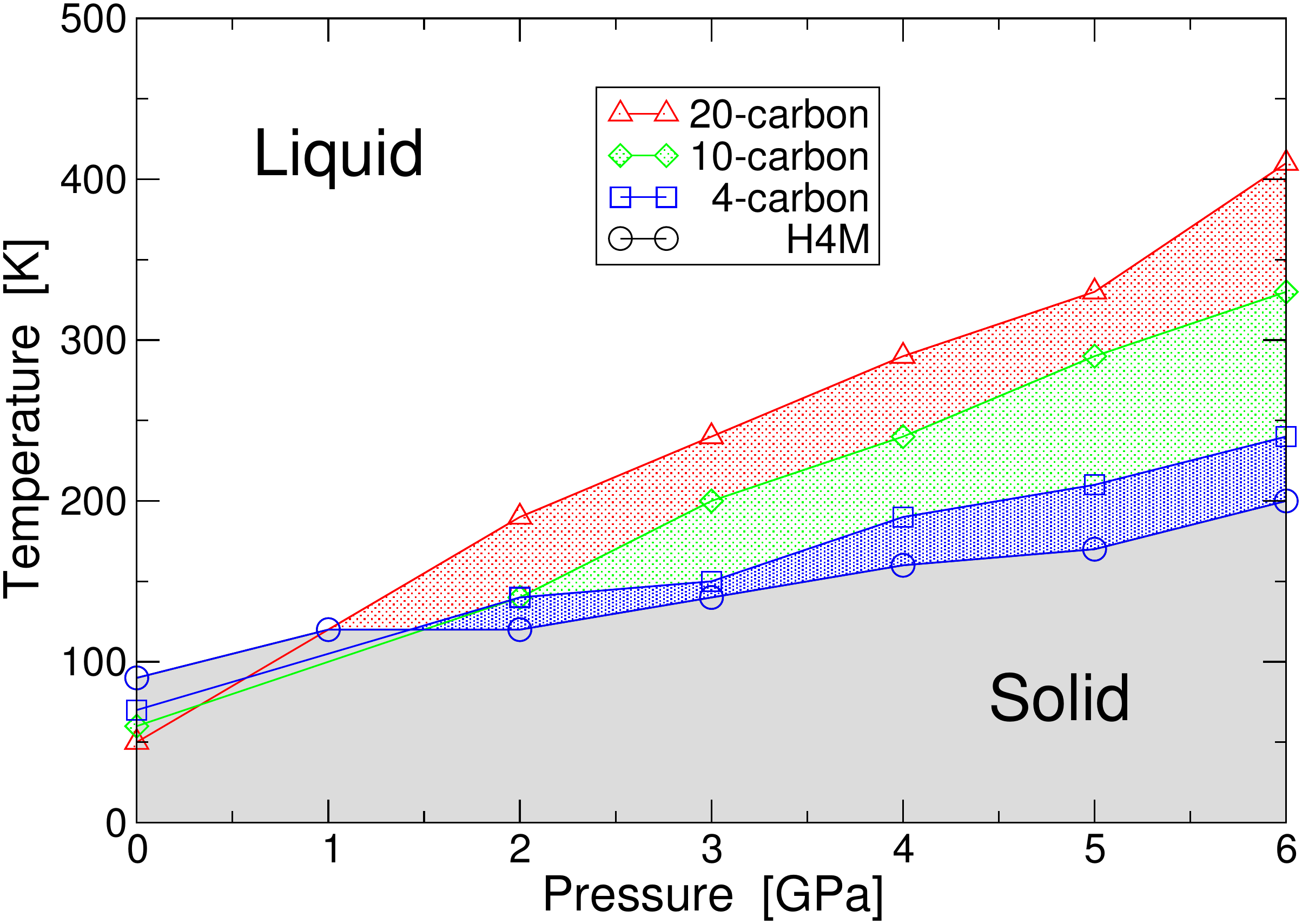}
\caption{\label{fig:phase}Calculated phase diagrams for several lengths
of H4-alkanes.}
\end{figure}

The most pertinent results of our study are collected in
Fig.~\ref{fig:phase}, where we depict the phase diagrams for the various
H4-alkanes. We do see a clear trend of increasing phase stability and
melting temperature as chain length increases. For the 20-carbon at
6~GPa the increase is over 200~K (relative to H4M), i.e.\ a doubling
of the stability field. From comparing the melting temperatures at lower pressures to those at
higher pressures, we see that, although there is a significant increase
in melting temperature compared to H4M for the longer chains at higher
pressures (i.e.\ 2--6~GPa), there is almost no difference or even a decrease in melting
temperature at lower pressures. To verify that this decrease is not merely an artifact of
the \textsc{Airebo} potential used, we also calculated the melting temperatures
for pure butane ($n=4$) and octane ($n=8$) at ambient pressures after performing an abbreviated structure search, finding melting temperatures
of 190~K and 290~K. Although these values are $\sim$60~K above the experimental values (see Fig.~\ref{fig:melting}), they still show
a consistent stabilization with chain length.
The most reasonable explanation for the decrease in stability at ambient pressure
is that there are competing interactions between the increased stability seen in the longer chains when they are aligned properly and the decreased stability from H$_2$ molecules serving as a barrier to intra-chain interactions. In order to benefit from the intra-chain interactions for the longer chains, they need to be oriented in a specific manner (i.e.\ aligned with each other as seen in Fig.~\ref{fig:structures}d). To maintain this orientation requires higher pressures, which is perhaps why these longer chains are less stable at ambient pressure. This case differs from that of regular alkanes due to the hydrogen molecules between the chains, which function as a barrier to intra-chain interactions.

As a side note, a 60-carbon structure was also studied, although due to
its size it was necessary to generate the structure randomly with
\textsc{Packmol}.\cite{Martinez_2009:packmol_package} No significant
improvement was found over the 20-carbon; analysis of the 60-carbon
structures near the end of the molecular dynamics simulations showed
significant chain breaking. It is possible that there is little
improvement in H4-alkane stability after a certain chain length due to
chain breaking. 
Also note that in Fig.~\ref{fig:phase}, except for H4M, there are no data points at
1~GPa. This is because for all of the structures other than H4M we could not obtain
complete and reliable MD trajectories. This most likely signifies an instability of
the structures at that particular pressure and possibly suggests a more complicated
phase transition.

Despite the significant improvement we see for the longer chains at high
pressures, the lack of stabilization at ambient pressures makes
H4-alkanes still not practical for automotive applications of hydrogen
storage, where even high-pressure tanks are typically limited to no more
than 0.1~GPa. Nonetheless, we see our work as a proof-of-principle that
longer chains significantly increase the stability field of H4-alkanes
and almost double it for the 20-carbon. Another point of importance is
that certain nonlinear alkane isomers, such as the cycloalkanes, have
higher melting temperatures than the straight alkanes. It is possible
that when H4-alkanes are created from these nonlinear alkanes, they
would have higher melting temperatures. However, for simplicity, as the
number of possible isomers tremendously grows with chain length, in this
work we restricted ourselves to linear alkane chains.
Furthermore, an almost limitless number of structures exist with only
slightly higher energy, which show intermingling of the
chains, which would further significantly increase the melting
temperature, but are unfortunately difficult to produce with a structure
search algorithm. 

Several other unexplored avenues also provide hope for
hydrogen storage applications of H4-alkanes (or derivative materials).
The simplest is modifying the H4-alkanes via adding stabilizing
molecules or doping it.  Another possibility is removing hydrogen from
the system, which has exceptional hydrogen content anyway and would not
drastically suffer from some hydrogen reduction. As hydrogen is removed,
H4-alkanes must approach the higher stability of pure alkanes (similarly, we would
expect the structures to become increasingly less stable as more than 4 H$_2$ molecules are added per carbon atom).  Assuming
the transition from the stability of H4-alkanes to regular alkanes
increases monotonically, it is likely that a class of materials such as
H2-alkanes (with 2 H$_2$ per carbon) would have significantly increased
stability. It is also worth mentioning that hydrogen can also be produced
through decomposing alkanes themselves.\cite{Otsuka_2002:production_hydrogen,Shah_2007:semi-continuous_hydrogen}
Finally, the overall phase diagram of the binary molecular system H$_2$ + CH$_4$ is very interesting,
but also highly complex---see Fig.~1 in Ref.~[\onlinecite{Somayazulu_1996:high-pressure_compounds}]. It is thus conceivable that other hydrogen-rich phases exist that may have more favorable stability
than H4-alkanes.

%%%%%%%%%%%%%%%%%%%%%%%%%%%%%%%%%%%%%%%%%%%%%%%%%%%%%%%%%%%%%%%%%%%%%%%%
\section{Conclusions}
%%%%%%%%%%%%%%%%%%%%%%%%%%%%%%%%%%%%%%%%%%%%%%%%%%%%%%%%%%%%%%%%%%%%%%%%

The material H4M has exceptional hydrogen storage density, but
unfortunately requires very high pressures and/or low temperature to be
stable. To overcome this shortcoming, we use the H4M structure and the
increasing stability of regular alkane chains as inspiration and predict
the possible formation of structures known as H4-alkanes (i.e.\ alkane
chains with 4 physisorbed H$_2$ molecules per carbon atom). We use a
structure search methodology to find candidate structures for H4-alkanes
of length 1, 4, 10, and 20 and our MD simulations show a significant
increase in stability for the longer chains at higher pressures. We see
our work as proof-of-principle that encourages further research to
stabilize H4-alkanes even more and thus realize their potential for
practical hydrogen storage applications.

%%%%%%%%%%%%%%%%%%%%%%%%%%%%%%%%%%%%%%%%%%%%%%%%%%%%%%%%%%%%%%%%%%%%%%%%
\begin{acknowledgements}
This work was supported in full by NSF Grant No.\ DMR-1145968.
\end{acknowledgements}
%%%%%%%%%%%%%%%%%%%%%%%%%%%%%%%%%%%%%%%%%%%%%%%%%%%%%%%%%%%%%%%%%%%%%%%%

%%%%%%%%%%%%%%%%%%%%%%%%%%%%%%%%%%%%%%%%%%%%%%%%%%%%%%%%%%%%%%%%%%%%%%%%
\bibliographystyle{apsrev4-1}
%\bibliography{final_refs}
%
%%%%%%%%%%%%%%%%%%%%%%%%%%%%%%%%%%%%%%%%%%%%%%%%%%%%%%%%%%%%%%%%%%%%%%%%

%%%%%%%%%%%%%%%%%%%%%%%%%%%%%%%%%%%%%%%%%%%%%%%%%%%%%%%%%%%%%%%%%%%%%%%%
\end{document}